\documentclass[11pt]{article}
\usepackage{amsmath}
\usepackage{amssymb}
\usepackage{amsfonts,amstext,latexsym,xspace}
\usepackage{epsfig}

\newcommand{\eq}{\begin{equation}}
\newcommand{\en}{\end{equation}}
\newcommand{\eqn}{\begin{eqnarray}}
\newcommand{\enn}{\end{eqnarray}}
\newcommand{\nn}{\nonumber }
\newcommand{\beq}{\begin{equation}}
\newcommand{\eeq}{\end{equation}}

\def\j2{$J_2^{{\bf C}}$}
\def\f4{$F_{4(4)}$}
\numberwithin{equation}{section}
\begin{document}
\begin{titlepage}
\begin{flushright}
  CERN-PH-TH/2006-107
\end{flushright}
\begin{center}
{\bf Orbits and Attractors for $N=2$ Maxwell-Einstein Supergravity
Theories in Five Dimensions}  \\

\vspace{1cm}
\begin{large}

 Sergio Ferrara$^{ \dagger}$\footnote{Sergio.Ferrara@cern.ch}
 and
 Murat G\"{u}naydin$^{\ddagger}$\footnote{murat@phys.psu.edu}
\end{large}
\\
\vspace{.35cm}

\vspace{.3cm} $^{\dagger}$ \emph{
Theory Division\\
CERN \\
CH-1211 Geneva 23, Switzerland \\
and \\
INFN  \\
Laboratori Nazionali di Frascati , \\
Via Enrico Fermi, 40, 00044, Frascati, Italy }\\
\vspace{.3cm} and \\
\vspace{.3cm}
$^{\ddagger}$ \emph{Physics Department \\
Pennsylvania State University\\
University Park, PA 16802, USA} \\

{\bf Abstract}
\end{center}

BPS and non-BPS orbits for extremal black-holes in $N=2$
Maxwell-Einstein  supergravity theories (MESGT) in five dimensions
were classified long ago by the present authors for the case of
symmetric scalar manifolds. Motivated by these results and some
recent work on non-supersymmetric attractors we show that attractor
equations in N=2 MESGTs in $d=5$ do indeed possess the distinct
families of solutions with finite Bekenstein-Hawking entropy. The
new non-BPS solutions have non-vanishing central charge and matter
charge which is invariant under the maximal compact subgroup
$\tilde{K}$ of the stabilizer $\tilde{H}$ of the non-BPS orbit. Our
analysis covers all symmetric space theories $G/H$ such that $G$ is
a symmetry of the action. These theories are in one-to-one
correspondence with (Euclidean) Jordan algebras of degree three. In
the particular case of $N=2$ MESGT with scalar manifold
$SU^*(6)/USp(6)$ a duality of the two solutions with regard  to
$N=2$ and $N=6$ supergravity is also considered.
\end{titlepage}

\section{Introduction}
\renewcommand{\theequation}{\arabic{section} - \arabic{equation}}
Extremal black hole solutions of  supergravity theories exhibit an
attractor mechanism" \cite{FKS} in their evolution towards the
horizon, in which the scalar fields move from their ( arbitrary)
asymptotic value $\phi_{\infty}$ toward a critical value
\[ r \rightarrow r_H \Rightarrow \phi(r) \rightarrow \phi_c(q) \]
which is determined by the critical points  of the  black hole
potential function \cite{FK,FGK} $V(\phi, q)$ such that  $\phi_c(q)$
is a solution of
\[ \partial_i V\equiv \frac{\partial}{\partial \phi^i } V=0 \]
The Bekenstein-Hawking area entropy $S$ is then given by
\begin{equation}
S \sim V\mid_{\partial_i V=0}
\end{equation}
in $d=4$ and
\begin{equation}
S\sim V^{3/4}\mid_{\partial_i V=0}
\end{equation}
in $d=5$.

For the case of scalar fields described by symmetric spaces of the
corresponding geometry of vector multiplets, the value of the
entropy is actually related to some invariants ( cubic in $d=5$ and
quartic in $d=4$) of the representation $R$ of the charge vector of
the black hole (B-H) charges \cite{KK,FM,FG,FK2}. For fixed
non-vanishing values of these invariants, the charge vectors
describe a $(dim R -1)$ dimensional orbits whose nature is strictly
related to the supersymmetry properties of the critical point. It
was pointed out in \cite{FG} that different orbits of charge vectors
correspond to different BPS and non-BPS configurations. Such non-BPS
configurations have been recently found in $d=4$ in some particular
cases \cite{GIJT,TT,K,KSS,SS,FGK,GJMT,G}, and this has prompted
further study in this direction. For example, in a recent work
\cite{FK2} it was shown that the $N=8$ attractors have in $d=4$ two
solutions,  of "maximal" symmetry, the $1/8$ BPS attractor with
$SU(2)\times SU(6)$ symmetry and the non-BPS attractor with $USp(8)$
symmetry. These symmetry groups are the maximal compact subgroups of
the stabilizers $E_{6(2)}$ and $E_{6(6)}$ of the two orbits
\footnote{We use the standard mathematical notation for labelling
non-compact real forms of Lie groups. The bracket in the subscript
is the difference between the number of non-compact generators and
compact generators.}
 \[
\frac{E_{7(7)}}{E_{6(2)}} \] and \[ \frac{E_{7(7)}}{E_{6(6)}}
\] found in \cite{FG}. They correspond to a positive and a negative
value of the quartic invariant $I_4$ in the 56 dimensional
representation of the charge vector of the $N=8$ theory. The
analysis of the four dimensional $N=2$ MESGTs and their BPS and
non-BPS orbits and attractors is treated in \cite{BFGM}.

In this paper we will study the five dimensional supergravity
theories. For the $N=8$ theory there is only one orbit with
non-vanishing entropy which is \cite{FG}
\[ \frac{E_{6(6)}}{F_{4(4)}} \]
The attractor nature of this orbit was derived in \cite{ADF} where a
solution of the attractor equation was shown to have $USp(6)\times
USp(2)$ symmetry which is the maximal compact subgroup of $F_{4(4)}$
\footnote{ We should note that the orbits of maximal supergravities
in $D$ dimensions were also studied in \cite{LPS}}. This corresponds
to a $1/8$ BPS attractor of the $N=8$ theory in $d=5$. Contrary to
the four dimensional case no other solution exists for the attractor
equation, in accordance with the analysis of \cite{FG}. However, in
\cite{FG}, it was shown that there exist two classes of orbits for
the $N=2$ MESGTs with symmetric scalar manifolds in $d=5$. These
orbits correspond to extremal BPS and non-BPS attractors in $d=5$.
In this paper we find the explicit solutions to the attractor
equations corresponding to these orbits.

In  section 2 we shall first review the real special geometry of
$N=2$ MESGTs as first formulated by  G\"unaydin, Sierra and Townsend
(GST). In section 3 we reproduce the classification of the orbits
with non-vanishing entropy using the theory of Jordan algebras. The
section 4 contains our main results on the solutions of the
attractor equations corresponding to the orbits classified in
\cite{FG}. In section 5 we discuss the scalar mass spectrum of the
solutions. In section 6 we consider the attractor equations for
self-dual strings in $d=6$ and in section 7 a concluding summary is
given.

\section{Geometry and symmetries  of $N=2$ Maxwell-Einstein supergravity
theories in five dimensions}

\renewcommand{\theequation}{\arabic{section} - \arabic{equation}}

In this section we will review symmetry groups of $N=2$ MESGT's in
five and four dimensions whose scalar manifolds  are symmetric
spaces. The MESGT's describe the coupling of an arbitrary number $n$
of (Abelian) vector multiplets to $N=2$ supergravity and five
dimensional MESGT's were constructed in \cite{GST}. The bosonic part
of the Lagrangian can be written as \cite{GST}
\begin{eqnarray}
   \label{Lagrange}
e^{-1}\mathcal{L}_{\rm bosonic}&=& -\frac{1}{2}R
-\frac{1}{4}{\stackrel{\circ}{a}}_{IJ}F_{\mu\nu}^{I}
F^{J\mu\nu}-\frac{1}{2}g_{xy}(\partial_{\mu}\varphi^{x})
(\partial^{\mu} \varphi^{y}) \nonumber \\ &&+
 \frac{e^{-1}}{6\sqrt{6}}C_{IJK}\varepsilon^{\mu\nu\rho\sigma\lambda}
 F_{\mu\nu}^{I}F_{\rho\sigma}^{J}A_{\lambda}^{K},
\end{eqnarray}
where $e$ and $R$ denote the f\"{u}nfbein determinant and the scalar
curvature in $d=5$, respectively. $F_{\mu\nu}^{I}$ are the field
strengths of the Abelian vector fields $A_{\mu}^{I}, \,( I=0,1,2
\cdots, n$) with $A^0_{\mu}$ denoting the ``bare'' graviphoton. The
metric, $g_{xy}$, of the scalar manifold $\mathcal{M}$ and the
``metric'' ${\stackrel{\circ}{a}}_{IJ}$ of the kinetic energy term
of the vector fields both depend on the scalar fields $\varphi^{x}$
( $x,y,..=1,2,..,n$). The n-dimensional  scalar manifold can be
identified with the  $\mathcal{V}=1$ hypersurface  of an $n+1$
dimensional ambient space  with coordinates $h^I$ and the metric
\begin{equation}\label{aij}
a_{IJ}(h):=-\frac{1}{3}\frac{\partial}{\partial h^{I}}
\frac{\partial}{\partial h^{J}} \ln \mathcal{V}(h) \ .
\end{equation}
where
\begin{equation}
\mathcal{V}(h):= C_{IJK} \, h^{I} h^{J} h^{K}\ .
\end{equation}
where $(I=0,1,\ldots,n)$.  We shall denote the flat indices on the
scalar manifold with lower case Latin indices $a,b,..=1,2,..,n$.

The metric  $g_{xy}$ of the scalar manifold is simply  the pull-back
of (\ref{aij}) to $\mathcal{M}$
\begin{equation}
  g_{xy} = h^I_x h^J_y {\stackrel{\circ}{a}}_{IJ}
\end{equation}
where
\begin{equation}
 h^I_x = -\sqrt{\frac{3}{2}} \frac{\partial}{\partial\phi^x} h^I
\end{equation}
and $ {\stackrel{\circ}{a}}_{IJ}$ is the ambient metric evaluated at
the hypersurface:
\begin{equation}
{\stackrel{\circ}{a}}_{IJ}\left(\varphi\right) = \left.
a_{IJ}\right|_{\mathcal{V}=1} \ .
\end{equation}
Supersymmetry implies further the algebraic constraints
\eqn
 {\stackrel{\circ}{a}}_{IJ}= h_I h_J + h_I^a h_J^a \nn \\
 h^Ih_I=1 \nn \\
 h^I_ah_I=h_I^a h^I=0 \\
 h^I_ah_b^J {\stackrel{\circ}{a}}_{IJ} = h^I_a h_{Ib}= \delta_{ab} \nn
 \enn
 as well as the differential constraints
 \eqn h_{I,x}= \beta h_{Ix} \nn \\
 h^I_{,x} = - \beta h^I_x \nn \\
 h_{Ix;y} = \beta ( g_{xy} h_I + T_{xyz} h^z_I ) \\
 h^I_{x;y} =-\beta (g_{xy} h^I + T_{xyz}h^{Iz} ) \nn
 \enn
where $\beta=\sqrt{\frac{2}{3}} $.

The Riemann curvature of the scalar manifold has the simple form
\begin{equation}
  K_{xyzu}= \frac{4}{3} \left( g_{x[u} g_{z]y} + {T_{x[u}}^w T_{z]yw} \right)
\end{equation}
where $T_{xyz}$ is  the symmetric tensor
\begin{equation}
   T_{xyz}= h^I_x h^J_y h^K_z C_{IJK}
\end{equation}
The full symmetry group of $N=2$ MESGT in $d=5$ is simply $ G \times
\mathrm{SU}(2)_R $ where $\mathrm{SU}(2)_R$ denotes the local
R-symmetry group of the $N=2$ supersymmetry algebra and $G$ denotes
the symmetry group of the tensor $C_{IJK}$. Now the covariant
constancy of $T_{xyz}$ implies the covariant constancy of
$K_{xyzu}$:
\begin{equation}
   T_{xyz; w} = 0 \Rightarrow  K_{xyzu ; w} =0
\end{equation}
Hence the scalar manifolds $\mathcal{M}_5$ with covariantly
 constant $T$ tensor are locally symmetric spaces.
If $\mathcal{M}_5$ is a homogeneous space the covariant constancy of
$T_{xyz}$ is equivalent to the following identity:
\begin{equation}
   C^{IJK} C_{J(MN} C_{PQ)K} = {\delta^I}_{(M} C_{NPQ)}
\end{equation}
where the indices are raised by ${\stackrel{\circ}{a}}{}^{IJ}$. For
proof of this equivalence an expression for constants $C_{IJK}$ in
terms of scalar field dependent quantities was used
\begin{equation}
   C_{IJK} = \frac{5}{2} h_I h_K h_K - \frac{3}{2} {\stackrel{\circ}{a}}_{(IJ} h_{K)} +
    T_{xyz} h^x_I h^y_J h^z_K
\end{equation}
as well as algebraic constraints $h_I h^I=1$ and $h^I_x h_I =0$ that
follow from supersymmetry \cite{GST}.   Using this  "adjoint
identity" , GST \cite{GST} proved that the cubic forms defined by
$C_{IJK}$ of $N=2$  MESGTs with symmetric target spaces
$\mathcal{M}_5$ ( with $n \geq 2$ ) and covariantly constant $T$
tensors are in one-to-one correspondence with the norm forms of
Euclidean (formally real) Jordan algebras $J$ of degree 3.
 The corresponding symmetric spaces are of the form
\begin{equation}
    \mathcal{M} = \frac{\mathrm{Str}_0 \left(J\right)}{ \mathrm{Aut}\left(J\right)}
\end{equation}
where $\mathrm{Str}_0\left(J\right)$ is the invariance group of the
norm (reduced structure group) and $\mathrm{Aut}\left(J\right)$ is
the automorphism group of the Jordan algebra $J$ respectively.

  Following  Schafers \cite{Schafers},
GST \cite{GST} listed the allowed cubic forms, which we reproduce
below:

\begin{enumerate}
\item $J=\mathbb{R}$  and   $\mathcal{V}(x) = x^3$.
 This
case corresponds to pure $d=5$ supergravity.
   \item $J = \mathbb{R} \oplus \Gamma$, where $\Gamma$ is a simple Jordan
   algebra with identity, which we denote as $\mathbf{e}_2$,   and
quadratic norm $Q\left(\mathbf{x}\right)$, for $\mathbf{x} \in
\Gamma$, such that $Q\left(\mathbf{e}_2\right)=1$.
 The norm is given as $\mathcal{V}\left(x\right) = a Q\left(\mathbf{x}\right)$, with
 $x=\left(a, \mathbf{x}\right) \in J$.
  This includes two special cases
\begin{enumerate}
  \item $\Gamma = \mathbb{R}$ and $Q = b^2$, with $\mathcal{V} = a b^2$. This is applicable to $n=1$.
  \item $\Gamma = \mathbb{R} \oplus  \mathbb{R}$ and $Q=bc$, and $\mathcal{V} = a b c$ and is applicable to $n=2$.
\end{enumerate}
For these special cases the norm is completely factorized, so that
$\mathcal{M}$ is flat. For $n>2$, $\mathcal{V}$ is still factorized
into a linear and quadratic parts. The positive definiteness of the
kinetic energy terms requires that $Q$ has Lorentzian signature
$\left(+,-,-,\ldots,-\right)$.   The invariance group of the norm is
 \begin{equation}
    \mathrm{Str}_0\left(J\right) = \mathrm{SO}\left(n-1, 1\right) \times \mathrm{SO}\left(1, 1\right)
 \end{equation}
where the $\mathrm{SO}\left(1, 1\right)$ factor arises from the
invariance of $\mathcal{V}$ under the dilatation $\left(a,
\mathbf{x}\right)
 \to \left( e^{-2\lambda} a, e^{\lambda} \mathbf{x}\right)$ for $\lambda \in \mathbb{R}$, and that
$\mathrm{SO}\left(n-1\right)$ is $\mathrm{Aut}\left(J\right)$. Hence
\begin{equation}
    \mathcal{M} = \frac{\mathrm{SO}\left(n-1,1\right)}{\mathrm{SO}\left(n-1\right)} \times \mathrm{SO}\left(1,1\right)
\end{equation}
This infinite family is referred to as the generic Jordan family of
MESGTs.

\item Simple Euclidean Jordan algebras  $J = J_3^\mathbb{A}$ generated by $3\times 3$ Hermitian matrices
over the four division algebras  $\mathbb{A}= \mathbb{R}$,
$\mathbb{C}$, $\mathbb{H}$, $\mathbb{O}$. An element $x \in
J_3^\mathbb{A}$ can be written as
\begin{equation}
   x = \begin{pmatrix}
         \alpha_1 &  a_3   & \bar{a}_2 \cr
         \bar{a}_3 & \alpha_2 & a_1 \cr
          a_2 & \bar{a}_1 & \alpha_3
     \end{pmatrix}
\end{equation}
where $\alpha_k \in \mathbb{R}$ and $a_k \in \mathbb{A}$ with
$\bar{}$ indicating the conjugation in the underlying division
algebra. The cubic norm $\mathcal{V}$ is given by
\begin{equation}
   \mathcal{V}\left(x\right) = \alpha_1\alpha_2\alpha_3 - \alpha_1 \left|{a_1}\right|^2 - \alpha_2 \left|{a_2}\right|^2
     - \alpha_3 \left|{a_3}\right|^2 + a_1 a_2 a_3 + \overline{\left( a_1 a_2 a_3
     \right)}
\end{equation}
 The corresponding spaces
$\mathcal{M}$ are irreducible of dimension $3 \left(1+\dim
\mathbb{A}\right)-1$, which we list below:
\begin{equation}
    \begin{array}{cc}
        \mathcal{M}( J_3^\mathbb{R}) =\phantom{gis }  & \dfrac{\mathrm{SL}\left(3, \mathbb{R}\right)}
        {\mathrm{SO}\left(3\right)} \\[12pt]
         \mathcal{M}(J_3^\mathbb{C} ) = \phantom{ges } & \dfrac{\mathrm{SL}\left(3, \mathbb{C}\right)}{\mathrm{SU}\left(3\right)}
    \end{array}
    \phantom{ and also }
    \begin{array}{cc}
         \mathcal{M}(J_3^\mathbb{H}) =\phantom{ges }  & \dfrac{\mathrm{SU}^\ast\left(6\right)}{\mathrm{USp}\left(6\right)} \\[12pt]
         \mathcal{M}(J_3^\mathbb{O}) =\phantom{ges } & \dfrac{\mathrm{E}_{6(-26)}}{\mathrm{F}_4}
    \end{array}
\end{equation}

\end{enumerate}


 The "magical" supergravity theories described by simple Jordan
algebras $J_3^\mathbb{A}$ ($\mathrm{A}$ = $\mathrm{R}$,
$\mathrm{C}$, $\mathrm{H}$ or $\mathrm{O}$) can be truncated to
theories belonging to the generic Jordan family. This is achieved by
restricting the elements of $J_3^\mathbb{A}$
\begin{equation}
    \begin{pmatrix}
         \alpha_1 &  a_3   & \overline{a}_2 \cr
         \overline{a}_3 & \alpha_2 & a_1 \cr
          a_2 & \overline{a}_1 & \alpha_3
    \end{pmatrix}
\end{equation}
to  their subalgebra $J = \mathbb{R} \oplus J_2^\mathbb{A}$ by
setting $a_1=a_2=0$. Their symmetry groups are as follows:
\begin{equation}
\begin{split}
    J = \mathbb{R} \oplus J_2^\mathbb{R}  &: \mathrm{SO}(1,1) \times \mathrm{SO}\left(2,1\right) \subset \mathrm{SL}\left(3, \mathbb{R}\right) \cr
    J = \mathbb{R} \oplus J_2^\mathbb{C}  &: \mathrm{SO}(1,1) \times \mathrm{SO}\left(3,1\right) \subset \mathrm{SL}\left(3, \mathbb{C}\right) \cr
     J = \mathbb{R} \oplus J_2^\mathbb{H}  &: \mathrm{SO}(1,1) \times \mathrm{SO}\left(5,1\right) \subset \mathrm{SU}^\ast\left(6\right) \cr
     J = \mathbb{R} \oplus J_2^\mathbb{O}  &: \mathrm{SO}(1,1) \times \mathrm{SO}\left(9,1\right) \subset \mathrm{E}_{6(-26)}
\end{split}
\end{equation}

\section{ Orbits of U-duality groups and Jordan algebras}
\renewcommand{\theequation}{\arabic{section} - \arabic{equation}}

Jordan algebras are commutative and non-associative algebras with a
 symmetric Jordan product $\circ$
\eq X \circ Y =  Y \circ X \en that satisfies the Jordan identity
\cite{jvw,nj} \eq X \circ ( Y \circ X^2 ) = (X \circ Y ) \circ X^2
\en Automorphism group of a Jordan algebra $J$ is formed by linear
transformations $A$ that preserve the products in $J$: \eq X\circ Y
= Z \Rightarrow (AX)\circ (AY) = (AZ) \en
The structure group
$Str(J)$ of $J$ is formed by linear transformations $S$ that
preserve the norm form $\mathcal{V}$ up to an overall scale factor
$\lambda$ : \eq Str(J) : X \rightarrow S(X) \Rightarrow
\mathcal{V}(S(X)) = \lambda \mathcal{V}(X) \en and the reduced
structure group $Str_0(J)$ of $J$ is the subgroup of $Str(J)$ that
leaves the norm  form invariant, i.e. those transformations $S$ for
which $\lambda=1$.

The Lie algebra $\mathfrak{aut}(J)$ of the automorphism group
$Aut(J)$  is generated by derivations $D$ that satisfy the Leibniz
rule:
\[ D(X\circ Y) = (DX)\circ Y + X \circ (DY) \]
It is easy to verify that by exponentiating derivations one obtains
automorphisms:
\[ e^{D} (X\circ Y) = (e^DX) \circ (e^D Y) \]
Hence it is customary to refer to $\mathfrak{aut}(J)$ as the
derivation algebra $Der(J)$ of $J$. Every derivation $D$ of $J$ can
be written in the form \cite{nj} \eq D_{X,Y} \equiv [ L_X,L_Y]
\hspace{1cm},  X,Y \in J \en where $L_X$ denotes multiplication by
$X$ i.e.
\[D_{X,Y} Z = X\circ (Y\circ Z)-Y\circ (X\circ Z) \]
Structure algebra $\mathfrak{str}(J) $ of $J$ is generated by
derivations and multiplication by elements of $J$:
\[ \mathfrak{str} (J)= Der(J) \oplus L_J \]
whose commutation relations are very simple \eqn [L_X,L_Y] = D_{X,Y}
\\ \nn [D_{X,Y},L_Z] = L_{(D_{X,Y}Z)} \\ \nn
[D_{X,Y}, D_{Z,W}] = D_{(D_{X,Y}W),Z} + D_{W,(D_{X,Y}Z)}   \enn
Multiplication by the identity element of $J$ commutes will all the
elements of $\mathfrak{str} (J)$ and acts like a central charge. The
reduced structure algebra $\mathfrak{str}_0 (J)$ is generated by
derivations and multiplications by traceless elements of $J$. The
automorphism group $Aut(J)$ leaves the identity element
$\mathfrak{e}$ of $J$ invariant or equivalently  derivations
annihilate the identity element \[ D\mathfrak{e}=0\]

As mentioned in the previous section, there exist four simple
(Euclidean) Jordan algebras of degree three $J_3^\mathbb{A}$ of
$3\times 3$ Hermitian matrices over the four division algebras
$\mathbb{A}=\mathbb{R}, \mathbb{C}, \mathbb{H}$ and $\mathbb{O}$. We
shall denote a general element $X$ of $J_3^\mathbb{A}$ as \eq X
=\left(
  \begin{array}{ccc}
    \alpha_1 & x_3 & \bar{x}_2  \\
    \bar{x}_3 & \alpha_2 & x_1 \\
    x_2& \bar{x}_1 & \alpha_3 \\
  \end{array}
\right) \equiv \sum_{i=1}^{3} \alpha_i E_i + (x_3)_{12} + (x_2)_{31}
+ (x_1)_{23} \en where $E_i$ ( $i=1,2,3$) are the 3 irreducible
idempotents of $J_3^{\mathbb{A}}$ and $x_i\in \mathbb{A}$ and the
bar denotes conjugation in $\mathbb{A}$. It is well-known that an
element $X$ of the algebra $J_3^{\mathbb{A}}$ can be diagonalized by
the action of the automorphism group. For the exceptional Jordan
algebra $J_3^\mathbb{O}$ this was shown explicitly in \cite{GPR}.
Thus \eq Aut(J_3^{\mathbb{A}}) : \hspace{1cm} X \Rightarrow (
\lambda_1 E_1 + \lambda_2 E_2 + \lambda_3 E_3 ) \en where
$\lambda_i$ are the eigenvalues of $X$. Norm of $X$   \[
\mathcal{V}(X)= \alpha_1 \alpha_2 \alpha_3 - \alpha_1 |x_1|^2
-\alpha_2 |x_2|^2 -\alpha_3 |x_3|^2 + 2 Re(x_1x_2x_3) \] is simply
the "determinant" and , hence, is equal to $\lambda_1 \lambda_2
\lambda_3 $.

Invariance group of the identity element $\mathfrak{1}$ is the
automorphism group $Aut(J)$. The subgroup of the automorphism group
that leaves an irreducible idempotent invariant is generated by
derivations that annihilate that idempotent. For , say, the
irreducible idempotent $E_3$ the corresponding derivations are: \eqn
D_{(x_3)_{12},(y_3)_{12}} \\ \nn D_{(x_3)_{12},(E_1-E_2)} \enn We
list , in Table 1,  simple Jordan algebras of degree three and
invariance groups $K$  of their irreducible idempotents that are
subgroups of  $Aut(J)$ .

\begin{table}
\begin{tabular}{|c|c|}  \hline
$J$ &$ K\subset Aut(J) $  \\ \hline $J_3^\mathbb{R}$ & $SO(2)\subset
SO(3)$ \\
$J_3^\mathbb{C}$   & $ SU(2)\times U(1) \subset SU(3) $ \\

$ J_3^\mathbb{H}$ & $ USp(4)\times USp(2) \subset USp(6) $ \\

$J_3^\mathbb{O}$  & $ SO(9) \subset F_4 $  \\ \hline
\end{tabular}
\caption{ Above we list the subgroups $K$ of the automorphism groups
of simple Euclidean Jordan algebras of degree three that leave an
irreducible idempotent invariant.}
\end{table}

The subgroup $K$ of the automorphism group that leaves the
idempotent $E_3$ invariant leaves also the element
\[ \mathfrak{b}= (-E_1-E_2+E_3) \]
with unit norm invariant. The little group of $\mathfrak{b}$ ,
defined as the subgroup of $Str_0(J)$ that leaves it invariant, is
generated by the above derivations plus the multiplications by the
traceless elements $(x_2)_{31}$  and $(x_1)_{23}$
\[ L_{(x_2)_{31}+(x_1)_{23}} = L_{(x_2)_{31}} + L_{(x_1)_{23}} \]
which are non-compact generators. The little groups $\tilde{H}$ of
$\mathfrak{b}$ for different Jordan algebras of degree 3 are listed
in Table 2.

\begin{table}
\begin{tabular}{|c|c|}  \hline
$J$ &$ \tilde{H} \subset Str_0(J) $  \\ \hline $J_3^\mathbb{R}$ &
$SO(2,1))\subset
SL(3,\mathbb{R})$ \\
$J_3^\mathbb{C}$   & $ SU(2,1) \subset SL(3,\mathbb{C}) $ \\

$ J_3^\mathbb{H}$ & $ USp(4,2) \subset SU^*(6)  $ \\

$J_3^\mathbb{O}$  & $ F_{4(-20)}  \subset E_{6(-26)} $  \\

$ \mathbb{R}\oplus \Gamma_n $ & $SO(n-2,1) \subset SO(n-1,1)\times
SO(1,1) $ \\ \hline
\end{tabular}
\caption{ Above we list the subgroups $\tilde{H} $ of the reduced
structure groups of  Euclidean Jordan algebras of degree three that
leave the element $\mathfrak{b}=-E_1 -E_2 + E_3$ invariant.}
\end{table}

As for the generic Jordan family $J= \mathbb{R}\oplus \Gamma_n $ ,
where $\Gamma_n$ is a Jordan algebra of degree two whose quadratic
norm has Lorentzian signature, we shall denote their elements in the
form:

\[ X= (\alpha; \beta_0, \vec{\beta}) \]
where \[ (\beta_0 \mathbf{1} + \vec{\beta}\cdot \vec{\sigma} ) \in
\Gamma_n \] The cubic norm of $X$ is simply
\eq \mathcal{V} = \alpha
( \beta_0^2 - \vec{\beta}\cdot \vec{\beta} ) \en
The three irreducible idempotents are \eqn E_1 = (1,0,\vec{0}) \nn \\
\nn E_2= ( 0; \frac{1}{2}, \frac{1}{2},0,...,0) \\
E_3 = ( 0; \frac{1}{2},-\frac{1}{2},0,..,0) \enn
with the identity
element $\mathfrak{1}$ given by
\[ \mathfrak{1} = E_1+E_2+E_3 = (1:1,\vec{0}) \]
Automorphism group of $J= \mathbb{R}\oplus \Gamma_n $ is $SO(n-1)$
and its reduced structure group is $SO(n-1,1)\times SO(1,1)$ . The
identity element $\mathfrak{1}$ is manifestly invariant under the
automorphism group and the little group of the element
$\mathfrak{b}=-E_1-E_2+E_3= (-1;0,-1,0,..,0)$ is $SO(n-2,1)$
\footnote{ Note that the invariance group of the idempotent $E_1$ is
$SO(n-1,1)$ , while the invariance group of the idempotents $E_2$
and $E_3$ is SO(n-2).}.

Let us now show that an element of $J_3^{\mathbb{A}}$ with positive
norm $\mathcal{V}(X)= \kappa^3 $ $(\kappa >0 )$ can be brought to
the form \eq\left(
              \begin{array}{ccc}
                \kappa & 0 & 0 \\
                0 & \kappa & 0 \\
                0 & 0 & \kappa\\
              \end{array}
            \right)
\en by the action of its reduced structure group if all its
eigenvalues are positive, or to the form
\eq\left(
              \begin{array}{ccc}
                -\kappa & 0 & 0 \\
                0 & -\kappa & 0 \\
                0 & 0 & \kappa\\
              \end{array}
            \right)
\en if two of its eigenvalues are negative. The global action of the
reduced structure group $Str_0(J)$ is generated by automorphisms and
by quadratic action $U_A$  by elements $A$ whose norm squared is
one. The quadratic operator $U_A$ acts on $J$ via: \eq U_A X
=\{AXA\} \equiv 2 (A\circ X)\circ A - A^2 \circ X \en and satisfies
the property  \eq \mathcal{V}(U_A X) = [\mathcal{V}(A)]^2
\mathcal{V}(X) \en Thus if $\mathcal{V}(A)=\pm 1$ then
\[ \mathcal{V}(U_A X)=\mathcal{V}(X) \]
Using the automorphism group one can bring an element $X$ to the
diagonal form :
 \eq  Aut(J) :
X \Rightarrow ( \lambda_1 E_1 + \lambda_2 E_2 +\lambda_3 E_3 ) \en
The quadratic action by $U_A$ that preserves the diagonal form must
involve $A$ which is also diagonal \[ A=A_1E_1+A_2E_2+A_3E_3 \] such
that $(A_1 A_2 A_3)^2=1$. Then \eq U_A (\lambda_1 E_1 + \lambda_2
E_2 +\lambda_3 E_3 ) \Rightarrow \lambda_1 A_1^2 E_1 + \lambda_2
A_2^2 E_2 +\lambda_3 A_3^2 E_3 \en This shows that one can rescale
the eigenvalues by a positive number such that the norm is
preserved. Hence an element with all positive eigenvalues can be
brought to a positive multiple of the identity $\kappa \mathfrak{1}$
where $\kappa^3=\lambda_1 \lambda_2 \lambda_3$. Consequently, the
orbit of a timelike (positive norm) element $X$ with all positive
eigenvalues is \eq \frac{Str_0(J)}{Aut(J)} \en
 The automorphism
groups of Jordan algebras of degree three and their reduced
structure groups are reproduced in Table 3.

\begin{table}
\begin{tabular}{|c|c|}  \hline
$J$ &$ $Aut(J)$  \subset Str_0(J) $  \\ \hline $J_3^\mathbb{R}$ &
$SO(3))\subset
SL(3,\mathbb{R})$ \\
$J_3^\mathbb{C}$   & $ SU(3) \subset SL(3,\mathbb{C}) $ \\

$ J_3^\mathbb{H}$ & $ USp(6) \subset SU^*(6)  $ \\

$J_3^\mathbb{O}$  & $ F_{4(-52)}  \subset E_{6(-26)} $  \\
$\mathbb{R}\oplus \Gamma_n$ & $SO(n-1) \subset SO(n-1,1)\times
SO(1,1) $ \\ \hline
\end{tabular}
\caption{ Above we list the subgroups $Aut(J) $ of the reduced
structure groups $Str_0(J)$  of Euclidean Jordan algebras of degree
three that leave the identity element  invariant.}
\end{table}

Similarly a timelike element with two negative eigenvalues and one
positive can be brought to the form ( modulo the permutation of the
diagonal entries)
\[ (-\kappa E_1 -\kappa E_2 + \kappa E_3 )  \]
where $\kappa^3 =\lambda_1\lambda_2\lambda_3 $ and the corresponding
orbit is
\[ Str_0(J)/ \tilde{H} \]
where $\tilde{H}$ is a non-compact real form of $Aut(J)$ listed in
table 2.

Extension to the generic Jordan family is straightforward and one
needs  only to use the standard knowledge of the orbits of the
Lorentz group in various dimensions. We should also note that orbits
with negative norm are isomorphic to the above orbits , depending on
whether all or one of the eigenvalues are negative.

\section{Attractor equations for $N=2$ MESGTs and their solutions in $d=5$ }
\renewcommand{\theequation}{\arabic{section} - \arabic{equation}}

We shall now consider the attractor mechanism in the framework of 5d
, $N=2$ MESGTs in an extremal B-H background described by a $(n+1)$
dimensional charge vector \[ q_I = \int_{S^3} H_I = \int_{S^3}
\stackrel{\circ}{a}_{IJ} *F^J \hspace{1cm} (I=0,1,...n) \] The
corresponding B-H potential , described in \cite{FK,FGK}, is
elegantly written in the framework of GST real special geometry as
follows:
\begin{equation}
V(\phi, q) = q_I {\stackrel{\circ}{a}}^{IJ} q_J
\end{equation}
where ${\stackrel{\circ}{a}}^{IJ}$ is the inverse of the metric
${\stackrel{\circ}{a}}_{IJ}$ of the kinetic energy term of the
vector fields. The metric ${\stackrel{\circ}{a}}_{IJ}$ is related to
the metric $g_{xy}$ of the scalar manifold via
\begin{equation}
{\stackrel{\circ}{a}}_{IJ} = h_I h_J + \frac{3}{2} h_{I,x} h_{J,y}
g^{xy} \label{eq:vectormetric}
\end{equation}
\[ {\stackrel{\circ}{a}}^{IJ} =h^Ih^J + \frac{3}{2} h^I_{,x}
h^J_{,y} g^{xy} \]
or conversely
\begin{equation}
g_{xy}= \frac{3}{2} h_{I,x} h_{J,y}{\stackrel{\circ}{a}}^{IJ} \en
Introducing  the quantity \[ Z= q_Ih^I \]
 we can write the
potential as \eq V(q,\phi) = Z^2 + \frac{3}{2} g^{xy} \partial_x Z
\partial_y Z \label{eq:N2} \en
where \[ \partial_x Z = q_I h^I_{,x} \]
 We should also note the identities
\eq T_{xyz}= (3/2)^{\frac{3}{2}} h^I_{,x} h^J_{,y} h^K_{,z} C_{IJK}
\en and \eq h^I_{,x;y} =\frac{2}{3}( g_{xy} h^I -\sqrt{\frac{3}{2}}
 T_{xyz} g^{zw} h^I_{,w} ) \en

The critical points of the potential are given by the solutions of
the equation \eq \partial_x V= 2 ( 2Z\partial_x Z - \sqrt{3/2}
T_{xyz} g^{yy'}g^{zz'} \partial_{y'}Z \partial_{z'} Z ) =0
\label{eq:derpotential}\en
For BPS critical points we have \eq
\partial_x Z=0 \en
and for non-BPS critical points \eq 2 Z \partial_x Z =
\sqrt{\frac{3}{2}} T_{xyz} \partial^y Z \partial^z Z
\label{eq:nonbps} \en where
\[ \partial^x Z \equiv g^{xx'} \partial_{x'} Z \]
The equation \ref{eq:nonbps} can be inverted using the relation
 \eq
q_I = h_I Z -\frac{3}{2} h_{I,x} \partial^x Z \en which follows from
equation \ref{eq:vectormetric}. For $\partial_x Z=0$ this gives \eq
q_I = h_I Z \en and for $\partial_x Z \neq 0$ we get \eq q_I = h_I Z
- (3/2)^{3/2} \frac{1}{2Z} h_{I, x} T^{xyz}
\partial_y Z \partial_z Z \en

Let us first remark that the attractor solution of the BPS orbit
with non-vanishing entropy  given by  \cite{FK2,CFGK} \[ \partial_x
Z=0 \] is invariant under the stability group $Aut(J)$ of the orbit
\[ Str_0(J)/Aut(J) \] listed in column 1 of table 1 of \cite{FG} ( see
table 3 above). If we now consider the second class of orbits
$G/\tilde{H}$ with non-vanishing entropy listed in column 2 of table
1 of \cite{FG} , we can solve the attractor equation by considering
$\partial_x Z$ invariant under the maximal compact subgroup $K$ of
$\tilde{H}$. The list of the stability groups $\tilde{H}$ and their
maximal compact subgroups $\tilde{K}$ is given in Table 4.

\begin{table}
\begin{tabular}{|c|c|c|}  \hline
$ J$ &$ {\tilde H}$& $ \tilde{K} $ \\ \hline $\mathbb{R}\oplus
\Gamma$
& $SO(n-2,1) $ & $SO(n-2)$ \\
$J_3^{\mathbb{R}}$ &$ SL(2,{\bf R})$ & SO(2)\\
$J_3^{\mathbb{C}}$ & $SU(2,1)$ & $SU(2)\times U(1)$\\
$J_3^{\mathbb{H}}$ &$ USp(4,2)$ & $USp(4)\times USp(2)$\\
$J_3^{\mathbb{O}}$ &$ F_{4(-20)} $& $SO(9)$ \\
 \hline
\end{tabular}
\caption{ Above we list the stability groups $\tilde{H}$ of the
non-BPS  orbits of the $N=2$ MESGT's with non-vanishing entropy in
$d=5$. The first column lists the Jordan algebras of degree 3 that
define these theories. The third column lists the maximal compact
subgroups $\tilde{K}$ of $\tilde{H}$.}
\end{table}

For the MESGTs defined by Jordan algebras of degree 3, the tensor
$C_{IJK}$ is an invariant tensor. Similarly the tensor $T_{abc}$ is
an invariant tensor of the maximal compact subgroup $H$. Going to
flat indices the attractor equation becomes: \eq 2Z \partial_a Z =
\sqrt{3/2} T_{abc} \partial^b Z \partial^c Z \label{eq:attractor}
\en The BPS solution is $\partial_a Z=0$ , which then gives \eq
V_{BPS}=Z^2 \en, using equation \ref{eq:N2}.  If $\partial_aZ \neq 0
$ , by  squaring  equation \ref{eq:attractor} we get \eq 4Z^2
\partial_a Z
\partial_a Z = \frac{3}{2} T_{abc} T_{ab'c'} \partial_b Z
\partial_c Z \partial_b' Z \partial_c' Z \en
Then using the identity
\[ T_{a(bc} T^a_{b'c')}= \frac{1}{2} g_{(bc} g_{b'c')} \]
valid only for the MESGTs defined by Jordan algebras of degree three
we get \eq \partial_a Z \partial_a Z = \frac{16}{3} Z^2 \en
 Hence at the non-BPS attractor point the potential becomes
 \eq
 V= Z^2 + \frac{3}{2} \partial_a Z \partial_a Z = 9 Z^2
 \label{eq:special}
 \en
 which agrees with the formula
 \eq
 V = (d-2)^2 Z^2
 \en
 valid for $d=4,5$ dimensions \cite{FK2,BFGM}.

Let us consider the example of the exceptional supergravity defined
by the exceptional Jordan algebra $J_3^{\mathbb{O}}$ whose U-duality
group is $E_{6(-26)}$ . The two extremal orbits are classified by
two different stabilizers , which are the compact $F_{4(-52)}$ and
the noncompact $F_{4(-20)}$ with the maximal compact subgroup
$SO(9)$. For the non-BPS orbit we have the decompositions \eqn
E_{6(-26)} \supset F_{4(-20)} \supset SO(9) \nn \\
27\rightarrow 26 +1  \nn \\
26=16+9+1 \enn Furthermore,
 \eq T_{abc} \Rightarrow (\bar{26})^3 = (\bar{16} \bar{16} \bar{9}) + (\bar{16} \bar{16} \bar{1} ) +
 (\bar{9} \bar{9} \bar{1}) +
 (\bar{1} \bar{1} \bar{1})
  \en It is easy to see that the solution \eq
\partial_a Z = ( \partial_{16} Z= \partial_9 Z=0 , \partial_1 Z \neq
0)\en  is a solution of equation \ref{eq:attractor} provided $ Z$
and $
\partial_1 Z$ satisfy the following algebraic equation \eq Z =
\frac{1}{2} \sqrt{\frac{3}{2}} T_{111} \partial_1 Z \en where we are
using flat coordinates. The entropy then becomes renormalized   as
in $d=4$  \cite{BFM} \eq S^{4/3} = V|_{\partial_{a} V=0} = Z^2 ( 1 +
\frac{4}{T_{111}^2} ) \en with the attractor point $T^2_{111}
=\frac{1}{2}$ for symmetric spaces defined by Jordan algebras.

 The scalar manifold
$\frac{SU^*(6)}{USp(6)}$ of the MESGT defined by $J_3^{\mathbb{H}}$
is the same as the scalar manifold of $N=6$ supergravity theory ,
whose attractor equation was studied in \cite{ADF}. Interestingly,
the $1/6$ BPS solution of $N=6$ supergravity in $d=5$ correspond
precisely to the orbit whose stabilizer is $USp(4,2)$ with maximal
compact subgroup $USp(4)\times USp(2)$. Indeed it was shown there
that the $6\times 6$ symplectic traceless matrix $Z^{AB}$ which
represent the the $N=2$ charges in the parent $N=6$ theory lead to
the following vacuum solution \eq Z^{AB} =\left(
          \begin{array}{ccc}
            e_1\epsilon & 0 & 0 \\
            0 & e_2\epsilon & 0 \\
            0 & 0 & e_3\epsilon \\
          \end{array}
        \right)
\en ,where $e_1+e_2+e_3=0$ and $\epsilon$ is the $2\times2$
symplectic matrix. At the attractor point we have  \[ e_2=e_3 \] i.e
$e_1=-2e_2$ at that point. The symplectic traceless matrix $Z^{AB}$
, which generically has a $USp(2)^3$ symmetry , has an enhancement
to $USp(2)\times USp(4)$ symmetry at the attractor point. This is
the non-BPS orbit of the $N=2$ subtheory , which is instead 1/6 BPS
in the $N=6$ theory. At the critical point the singlet $X$ of the
$N=6$ supergravity takes the value \[ e_1=\frac{8}{3} X \] and the
entropy \footnote{To compare with the formula of $5d$ $N=2$ geometry
one should take into account the normalization of the terms in
\ref{eq:N2} compared to \ref{eq:N6}.}

\eq V=\frac{1}{2} Z^{AB}Z_{AB} + \frac{4}{3} X^2  \label{eq:N6} \en
becomes \eq V=e_1^2 + e_2^2 + (e_1+e_2)^2 +\frac{4}{3} X^2 = 12 X^2
\en Therefore at the attractor point \eq V_{NBPS}(X) = 12 X_{NBPS}^2
\en while at the BPS attractor point \eq V_{BPS}=\frac{4}{3}
X_{BPS}^2 \en It is easy to check this result with the cubic
invariant as given in \cite{ADF} \eq I_3= -\frac{1}{6} Tr (ZC)^3
-\frac{1}{6} Tr(ZC)^2 X + \frac{8}{27} X^3 \en where $C$ is the
$USp(6)$ invariant symplectic metric. At the BPS attractor point
$Z=0$ \eq I_3 =\frac{8}{27} X^3 \en with the entropy \eq S^{4/3}
=V|_{BPS} = 3 |I_{3}|^{2/3}=\frac{4}{3} X_{BPS}^2 \en and at the
non-BPS point \eq -I_3 = \frac{216}{27} X^3 = 8 X_{NBPS}^3 \en with
the entropy  \[ S^{4/3}= 3|I|^{2/3} = 12 X_{NBPS}^2
\]
 The $N=2$
derivation of the above result is through formula \ref{eq:special}
of the real special geometry of Gunaydin, Sierra and Townsend.
\section{Scalar masses at the attractor points}
\renewcommand{\theequation}{\arabic{section} - \arabic{equation}}
One can give general results on the quadratic fluctuations of the
B-H potential $V$ around its BPS and non-BPS critical points. The
general form of $V$ is given in equation \ref{eq:N2} . By further
differentiating equation \ref{eq:derpotential}  we get a general
expression for the Hessian of the potential: \eqn \frac{1}{4} D_x
\partial_y V &=& \frac{2}{3} g_{xy} Z^2 + \partial_x Z \partial_y Z
-2\sqrt{\frac{2}{3}} T_{xyz} g^{zw} \partial_w Z Z \label{eq:masses} \\
&& +
T_{xpq}T_{yzs} g^{pz} g^{qq'}g^{ss'}\partial_{q'} Z \partial_{s'} Z \nn  \\
&=& \frac{2}{3}( g_{xz} Z -\sqrt{\frac{3}{2}} T_{xzp} \partial^p Z )
( g_{yz}Z - \sqrt{\frac{3}{2}} T_{yzq} \partial^q Z ) + \partial_x Z
\partial_y Z  \nn \enn
From the above equation we obtain the Hessian at the BPS critical
point $\partial_xZ=0$ \eq \partial_x \partial_y V =\frac{8}{3}
g_{xy} Z^2 \label{eq:positivemass} \en which is the same result as
in $d=4$ \cite{FGK}. Note that equation \ref{eq:positivemass}
implies that the scalar fluctuations have positive square mass which
shows the attractor nature of the BPS critical points
\cite{FGK,GIJT}-\cite{G}. At the non-BPS critical point we can split
the index $x=(p,1)$ where $1$ is the singlet direction along the
subgroup $K$ of the stabilizer of the orbit. The using flat
coordinates the attractor condition \ref{eq:derpotential} becomes
\eqn \partial_1 Z = \frac{4}{\sqrt{3}} Z \nn
\\
\partial_pZ=0 \\
(T_{111}=\frac{1}{\sqrt{2}} ) \nn \enn which when inserted in
\ref{eq:masses} gives \eq \frac{1}{4} \partial_p\partial_q V =
\frac{2}{3} ( \mathbb{I} - T)^2_{pq} Z^2 = \frac{2}{27} (
\mathbb{I}-T)_{pq}^2 V_{NBPS} \label{eq:critical} \en where
 $({\mathbb{I}} -
T)_{pq} = \delta_{pq} -T_{pq} $ and $T_{pq}=2\sqrt{2} T_{pq1} $. For
the singlet mode , $T_{111}=\frac{1}{\sqrt{2}}$ and then  \eq
\frac{1}{4} V_{11}= (\frac{2}{3}+\frac{16}{3})Z^2 = 6Z^2
=\frac{2}{3} V_{NBPS} \en

We can further split the indices $(p,q)$ into $(i,j),  (\alpha,
\beta )$ where $(i,\alpha)$ refer to the two representations with
non-vanishing T-tensor given by \eq T_{\alpha \beta 1}, T_{ij1},
T_{\alpha \beta i} \en From the identities \eq T^1_{(\alpha \beta}
T^1_{\gamma\delta)} + T^i_{(\alpha \beta} T^i_{\gamma\delta)}
=\frac{1}{2} \delta_{(\alpha \beta} \delta_{\gamma\delta)}
\label{eq:tt} \en

\eq T^1_{(ij}T^1_{lm)} =\frac{1}{2}\delta_{(ij} \delta_{lm)} \en We
have $T^1_{ij}=\frac{1}{\sqrt{2}} \delta_{ij}$  and thus \eq
\frac{1}{4} \partial_i \partial_j V =\frac{2}{27} V_{NBPS}
\delta_{ij} \en
From \ref{eq:tt} we also have \footnote{ Here we
used the identity $\gamma_{\mu (ij} \gamma^\mu_{kl)}= \delta_{(ij}
\delta_{kl)} $ which follows from the fact that $SO(9)$ Clifford
algebra is isomorphic to $J_2^{\mathbb{O}}$, and which can also be
proven using more traditional methods \cite{DVV}.}

\eq T^1_{\alpha\beta} = \lambda \delta_{\alpha \beta}\en  with \eq 0
<\lambda <\frac{1}{\sqrt{2}} \en If $\lambda=\frac{1}{2\sqrt{2}}$ so
that $T_{\alpha\beta} =\delta_{\alpha\beta} $ we have \eq
\partial_{\alpha}\partial_{\beta} V=0 \en
For the $J_3^H$ model this would agree with the splitting of the 14
scalars into (5+1) massive vector multiplets and 2 massless
hypermultiplets according to the $N=6$ interpretation of its moduli
space. Equation \ref{eq:critical} implies that for the non-BPS
critical points the scalar square-mass matrix is semi-positive
definite. For the massless fluctuations,   attractor nature of the
solution depends on  third or higher derivatives \cite{TT}-\cite{G}.
We leave this problem to a future investigation.
\section{ Attractors in six dimensions}
\renewcommand{\theequation}{\arabic{section} - \arabic{equation}}
The analysis of the previous section can be extended to six
dimensions which is the maximal dimension where theories with 8
supercharges exist. These are the (1,0) theories describing the
coupling of $n_T$ tensor multiplets to supergravity
\cite{Romans,Sagnotti}. In this case, as discussed in \cite{DFKR},
the string tension plays the role of central charge and it depends
on the tensor multiplet scalars through the coset representative
$X_I$ of the
\[ \frac{SO(1,n_T)}{SO(n_T)} \] $ \sigma $ model \cite{Romans,Sagnotti} \eqn
Z=X^{\Lambda}q_{\Lambda} \nn \\ X^{\Lambda}\eta_{\Lambda \Sigma}
X^{\Sigma} =1 \enn
 where $\eta_{\Lambda \Sigma}$ is the $(1,n_T)$
Lorentzian metric $(\Lambda, \Sigma = 0,1,..,n)$. The matter charges
are $X_{\Lambda I}$ ($I=1,...,n_T)$ with the
property \eqn X_{\Lambda} X_{\Sigma} -X_{I \Lambda } X_{I \Sigma } = \eta_{\Lambda \Sigma} \\
X_{\Lambda} X_{\Sigma}  + X_{I \Lambda } X_{I \Sigma } =
\mathcal{N}_{\Lambda \Sigma} = 2X_{\Lambda} X_{\Sigma} -
\eta_{\Lambda \Sigma} \nn \enn and $\mathcal{N}_{\Lambda \Sigma}$ is
the metric of the kinetic energy of the self-dual tensor fields. As
shown in \cite{ADFL} if one defines a black string  potential energy
as \eq V= q^{\Lambda} \mathcal{N}_{\Lambda \Sigma} q^{\Sigma} =
q_{\Lambda}  \mathcal{N}^{\Lambda \Sigma  } q_{\Sigma} = Z^2 + Z_I^2
\en  where $Z_I = X_{I\Lambda} q^{\Lambda} $ and the inverse formula
, analogue of 4-10,  holds \[ q^{\Lambda} = X^{\Lambda} Z -
X_I^{\Lambda} Z_I \] it is easy to show that the attractor condition
$\partial_IV=0$ implies \cite{ADFL} \eq ZZ_I=0 \en with the solution
$Z \neq 0$ , $Z_I=0$ which is the BPS case and the other , $Z_I\neq
0$, $Z=0$ , is the non-BPS case and corresponds to "tensionless"
strings. However, in both cases the string energy is non-vanishing
and, moreover,  the two cases correspond to time-like and space-like
configurations of the charges since \eq q^{\Lambda} \eta_{\Lambda
\Sigma}q^{\Sigma} = Z^2 - Z_I^2 \en is an invariant. Therefore , as
expected, the two classes of orbits are \eq
\frac{SO(1,n_T)}{SO(n_T)} \hspace{2cm} BPS \en \eq
\frac{SO(1,n_T)}{SO(1,n_T-1)} \hspace{2cm} non-BPS \en

\section{Conclusions} In this paper we have examined the nature of
attractor equations for 5D extremal black holes based on the
geometry $N=2$  vector multiplet scalars which are symmetric spaces
$G/H$ such that $G$ is a symmetry of the MESGT \cite{GST}.

There are two generic classes of attractors, one BPS and one
non-BPS, described by two classes of charge orbits previously found
by the same authors. The value of the B-H potential at the non-BPS
attractor point is \[ V|_{\partial_x V=0}= (3Z_{NBPS})^2 =
|I_3|^{2/3}
\]
so that the Bekenstein-Hawking entropy is given, in terms of the
central charge,  by
\[ S_{NBPS} = (3Z_{NBPS})^{3/2} = \sqrt{|I_3|} \]

This is to be compared with the supersymmetric attractors  where
\cite{FK2,ADF,CFGK}
\[ V|_{\partial_x  V =0}=Z^2 = I_3^{2/3} \]
\[ S_{BPS} = (Z)^{3/2}_{BPS} = \sqrt{I_3} \]
 For non-symmetric spaces the analysis may change because the
 derivation of equation $V=9Z^2$ will no longer be valid. It would
 be interesting to consider the case of Calabi-Yau compactifications
 \cite{CCDF} in which case the $C_{IJK}$ coefficients are related to
 the C-Y intersection numbers. From the general validity of the
 equations 4-50-4.53, in the one modulus case, the renormalized
 formula \[ S^{4/3}=V|_{\partial V=0} = Z^2 ( 1 + \frac{4}{T_{111}^2}) \]
 will still hold in analogy with a similar situation in $d=4$
 \cite{BFM}.

 We have also analyzed the black self-dual string potential in the
 case of $d=6$ $(1,0)$ theories.  In that case, due to the nature
 of the tensor moduli space, there are just two orbits with
 non-vanishing entropy, a time-like one corresponding to BPS
 attractors and a space-like one corresponding to non-BPS attractors
 which are tensionless strings ( zero central charge). The $d=4$
 situation is much richer and will be considered elsewhere \cite{BFGM}.

{\bf Acknowledgement:}
 This work was
supported in part by the National Science Foundation under grant
number PHY-0245337 and PHY-0555605. Any opinions, findings and
conclusions or recommendations expressed in this material are those
of the authors and do not necessarily reflect the views of the
National Science Foundation. The work of S.F. has been supported in
part by the European Community Human Potential Program under
contract MRTN-CT-2004-005104 " Constituents, fundamental forces and
symmetries of the universe", in association with INFN Frascati
National Laboratories and by the D.O.E grant DE-FG03-91ER40662, Task
C.

\end{document}